\magnification=1200
\baselineskip14pt plus1pt minus1pt
\overfullrule0pt

\font\bff=cmb10
\font\bb=msbm10

\def\R{\hbox{\bb R}}
\def\bs{\bigskip}
\def\ms{\medskip}
\def\ss{\smallskip}
\def\nt{\noindent}
\def\dis{\displaystyle}

\def\i#1/#2\par{\item{\hbox to .6truecm{#1\hss}}#2}
\def\ii#1/#2/#3\par{\advance\parindent by .8truecm
       \item{\hbox to .6truecm{#1\hss}
         \hbox to 0.6truecm{#2\hss}}#3
          \advance\parindent by -.8truecm}
\def\iii#1/#2/#3/#4\par{\advance\parindent by 1.6truecm
        \item{\hbox to 0.6truecm{#1\hss}
              \hbox to 0.6truecm{#2\hss}
              \hbox to 0.6truecm{#3\hss}}#4
        \advance\parindent by -1.6truecm}


\centerline{\bff THE TOPOLOGY OF THE AdS/CFT/RANDALL-SUNDRUM} 
\centerline{\bff COMPLEMENTARITY}
\bs
\centerline{Brett McInnes}
\ms
\centerline{Department of Mathematics}
\ss
\centerline{National University of Singapore}
\ss
\centerline{10 Kent Ridge Crescent, Singapore 119260}
\ss
\centerline{Republic of Singapore}
\ss
\centerline{email: matmcinn@nus.edu.sg}
\bs\bs\bs

\nt{\bf ABSTRACT:} \ \ The background geometries of the AdS/CFT and the Randall-Sundrum theories are {\it locally} similar, and there is strong evidence for some kind of ``complementarity'' between them; yet the {\it global} structures of the respective manifolds are very different.  We show that this apparent problem can be understood in the context of a new and more complete global formulation of AdS/CFT.  In this picture, the brane-world arises {\it within} the AdS/CFT geometry as the inevitable consequence of recent results on the global structure of manifolds with ``infinities''.  We argue that the usual coordinates give a misleading picture of this global structure, much as Schwarzschild coordinates conceal the global form of Kruskal-Szekeres space.
\bs

\nt PACS : 11.25.-w, 11.25.Hf
\ss
\nt {\bf Keywords : Randall-Sundrum, AdS/CFT, Euclidean Quantum Gravity}
\ss
\nt LANL Database Number : hep-th/0009087 
\bs

\nt {\bf I. \ INTRODUCTION.}
\ss
In the best-known special case, the AdS/CFT correspondence [1] is a duality between quantum gravity on an $(n+1)-$dimensional AdS background (for certain $n$) and a Yang-Mills superconformal field theory on the $n-$dimensional boundary of the Penrose compactification.  That this is indeed a {\it special} case is emphasised in Ref. [2], where it is also argued that more general cases are best treated by Euclidean methods.  In the Euclidean picture, the bulk is an $(n+1)-$dimensional hyperbolic space $H^{n+1}$ (topology $\R^{n+1})$; it may be regarded as the warped product of $\R$ with $\R^n$, with metric
$$g^H = dy \otimes dy + e^{+2y/L} \; g^F_{ij} \; dx^i \otimes dx^j, \eqno{({\rm I}.1)}$$
\nt where $y \in (-\infty, \infty)$, $L$ is a constant, $g^F_{ij} \; dx^i \otimes dx^j$ is a flat Euclidean metric, and the Ricci curvature of $g^H$ satisfies
$${\rm Ric}(g^H) = -{n\over L^2} \; g^H. \eqno{({\rm I}.2)}$$
\nt The conformal boundary consists of a copy of $\R^n$ at ``$y=+\infty$'' together with a point at ``$y=-\infty$'', and so it has the topology of $S^n$, the $n-$sphere, the one-point compactification of $\R^n$.  [Notice that the sections $y$ = constant have a different topology : each is a copy of $\R^n$.]  The metric $g^H$ induces a conformal structure rather than a metric on the boundary.  This is the purely geometric aspect of the duality.

\ms

The AdS/CFT correspondence has proved to be an extremely useful tool for exploring otherwise inaccessible aspects of quantum gravity.  It is natural to ask whether it is more than this : {\it can something resembling} [a Euclidean version of] {\it the real world arise within the} AdS/CFT {\it geometry}?  One immediately thinks, of course, of the Randall-Sundrum ``brane world'' [3], which sits at the origin of a 5-dimensional space with metric
$$g^{RS} = dy \otimes dy + e^{-2|y|/L} \; g^F_{ij} \; dx^i \otimes dx^j. \eqno{({\rm I}.3)}$$

\nt Indeed, evidence for some kind of AdS/CFT/Randall-Sundrum ``complementarity'' has been adduced [4] [5] [6] [7] [8] [9] [10] [11] [12] [13] [14] [15] [16].  For example, in Ref. [10] the corrections to the Newtonian gravitational potential are computed from both points of view, and are found to have the same form.  The agreement becomes precise {\it provided} that the curvature parameter $L$ in the Randall-Sundrum metric is equated to its counterpart in $g^H-$ again strongly suggesting that we are dealing here with two aspects of one geometry.  A clearer and more concrete interpretation would be that we are dealing with two different {\it regions} of a single Riemannian manifold with curvature parameter $L$.

\ms

While $g^H$ and $g^{RS}$ are obviously closely related, the global structures of the respective manifolds are of course very different; and these differences are crucial.  First, the Randall-Sundrum space has no access to any region analogous to the region of $H^5$ where $y$ takes large positive values, and this is essential for ``trapping'' gravity on the brane-world.  Second, this excision of the ``large end'' of $H^5$ results in a manifold which is not geodesically complete.  Again, this incompleteness (due to the ``jump'' at $y=0$) may well play an essential role in gravity trapping (see [14] [16], and the references therein).  Thirdly, as was pointed out above, the topology of the conformal boundary of $H^5$ is that of $S^4$, and thus is very different to that of the brane-world; this is no mere technical bagatelle, for the compactness of the boundary ensures the uniqueness of the correlation functions of the conformal field theory defined there.

\ms

Despite these global differences, there is no denying the appeal of the notion that the brane world results from a $UV$ cutoff of AdS/CFT, with a corresponding non-trivial coupling of the CFT to gravity [4].  From this point of view, AdS/CFT should be regained from the Randall-Sundrum space by some kind of limiting process.  But what is this process?  How can it resolve the global differences discussed above?  The metrics $g^H$ and $g^{RS}$ appear to have no adjustable parameter other than $L$; how can we introduce a new parameter into $g^{RS}$, such that $g^H$ is obtained in the limit?  What is the geometric significance of this new parameter?

\ms

We wish to argue that these questions are best answered in the context of a radical reformulation of the geometric background of AdS/CFT.  The fact that the familiar bulk/boundary formulation of AdS/CFT must be generalised in some way is already recognised in [2] : ultimately, we expect a fully
general AdS/CFT correspondence to relate gauge theory on an {\it arbitrary}
compact manifold to string theory on a related higher-dimensional Einstein
manifold of negative scalar curvature. However, as we move beyond simple manifolds like $S^n$ and $S^1 \times S^{n-1}$, we soon encounter manifolds of non-zero Hirzebruch signature, and these simply cannot be represented as the boundary of {\it any} manifold-with-boundary [17].  A familiar example of this is the $K3$ manifold. Even if we leave this problem to one side,
there are serious practical difficulties in carrying out this programme.
For example, every four-dimensional compact flat manifold is the quotient
of the torus $T^4$ by some finite but possibly very complicated group. All
such manifolds can be represented as the boundary of some five-dimensional
manifold-with-boundary, but this fact is established in an abstract way
using index theory, not by a concrete construction; to exhibit an Einstein
metric on the interior of this space [inducing on the boundary the 
conformal structure corresponding to the given flat metric] is, in general,
a formidable task indeed. We have shown elsewhere [18] that the first of
these problems is solved by interpreting ``infinity'' as a compact submanifold of a compact {\it manifold} rather than as the boundary of a manifold-with-boundary. [As we shall explain below, cobordism theory
indicates that this is really the only natural way of dealing with this
problem.] The ``bulk'' is then the (non-compact) complement of the infinity hypersurface.  This is indeed a generalisation of the usual approach; for while it is {\it essential} if we are to deal with non-boundary manifolds like $K3$, it also works perfectly for manifolds like spheres and tori which happen to be boundaries. The second problem is then automatically
solved, since, as we shall show here, it is quite easy to exhibit Einstein
metrics defined on the bulk if the latter is interpreted in this way.

\ms

If we accept that this is the correct geometric formulation of AdS/CFT, then a pair of beautiful results due to Witten and Yau [19] and Cai and Galloway [20] essentially imply the following remarkable result : if we require the bulk to be Einstein with negative scalar curvature, then the bulk {\it cannot be geodesically complete}.  Something like the brane-world arises in the bulk not because it has been inserted ``by hand'', {\it but because the geometry requires it to do so}.  The Einstein condition on the bulk strongly constrains the {\it local} geometry, and yet, by forcing us (via incompleteness) to ``cut and paste'', it opens up a host of interesting new global structures for the higher-dimensional space.  A very simple and natural example of this construction will allow us to represent the brane-world as residing in a space which is essentially connected to the ``AdS/CFT region'' by a wormhole.  The solution contains, as so often, negative-tension branes, but these are confined to the throats of the wormhole - where, one might say [21], they belong.  The parameter which controls the extent to which the brane world resembles AdS/CFT infinity is the width of the wormhole, or, equivalently, the tension of the wormhole branes; the topology of infinity is always identical to that of the brane world; and there is a clear and precise sense in which the AdS/CFT picture is obtained from the brane world as the parameter tends to its limiting value. Finally, the wormhole approach can be reconciled with the usual
bulk/boundary formulation of AdS/CFT by taking another limit of this
parameter, in which the wormhole effectively pinches off and converts
the infinity hypersurface to an effective boundary, destroying all the
branes and restoring geodesic completeness. Thus, the new approach
does not completely replace the standard one: instead it reveals the 
latter as a special, limiting case of a more general geometry in which
the bulk is generically non-smooth. In short, the reformulation advocated here allows us to understand the differences between the AdS/CFT picture and the Randall-Sundrum cosmology, and provides a suitable arena for discussing their relationship.

\ms

We begin with a rapid summary of the relevant geometric ideas.

\bs

\nt {\bf II. \ SPIN COBORDISM AND EINSTEIN METRICS}
\ss
Let $N^n$ be a smooth, compact, $n-$dimensional submanifold of a compact $(n+1)-$  dimensional manifold ${\hat M}^{n+1}$.  Let $f$ be a smooth function on ${\hat M}^{n+1}$ which is positive on the complement $M^{n+1} = {\hat M}^{n+1} - N^n$ but which vanishes to first order on $N^n$.  Suppose that we can find a metric $g^M$ on $M^{n+1}$ and a function $f$ as above, such that $f^2\; g^M$ extends continuously to $N^n$.  Then $N^n$ is ``at infinity'' for points in the ``bulk'', $M^{n+1}$.  Any instance of the usual bulk/boundary formulation of AdS/CFT can be interpreted in this way by simply taking two copies and identifying them along the boundary.  A more interesting construction results, however, when one considers how to deal with a CFT defined on a manifold which is {\it not} a boundary.

\ms

Let $P^n$ and $Q^n$ be compact $n-$dimensional spin manifolds.  Let $-Q^n$ be obtained from $Q^n$ by reversing orientat
ion, and let $P^n + (-Q^n)$ be the disjoint union.  One says that $P^n$ and $Q^n$ are {\it spin cobordant} [22] if there exists at least one compact manifold-with-boundary ${\bar R}^{n+1}$ with $P^n + (-Q^n)$ as boundary and with an interior $R^{n+1}$ which is spin and which induces the given spin structures on $P^n$ and $-Q^n$.  The spin cobordism equivalence classes in dimension $n$ form an abelian group, the spin cobordism group $\Omega_n^{\rm{spin}}$.  In particular, $\Omega_4^{\rm{spin}}$ is not trivial : there are (infinitely) many distinct 4-manifolds which are not spin boundaries.

\ms

Let $N^n$ be a compact manifold on which we wish to investigate the
generalised AdS/CFT correspondence [2], but suppose that $N^n$ is not
spin cobordant to zero, so it is not the spin boundary of any manifold-
with-boundary. The fact that $\Omega_n^{\rm{spin}}$ is a {\it group} points
the way to a solution of the problem, for clearly the class represented by
$N^n$ has an inverse: that is, the non-boundary $N^n$ can always be 
cancelled by another non-boundary, say $P^n$. That is, there exists a 
manifold-with-boundary ${\bar M}^{n+1}$ with $N^n + (-P^n)$ as boundary.
Let $g^M$ be a metric on the interior, $M^{n+1}$, such that $N^n + (-P^n)$
is the conformal boundary relative to $g^M$; thus we have represented $N^n$
as one connected {\it component} of the boundary, which is the best we
can do. However, there are strong objections to disconnected boundaries
in the generalised AdS/CFT correspondence [19], so this procedure only
exchanges one problem for another. The only way to avoid this conclusion
is to pick $P^n$ to be $N^n$ itself, and then perform a topological 
identification, obtaining a compact manifold ${\hat M}^{n+1}$ ({\it without} a boundary) which contains $N^n$ as a ``hypersurface at infinity''. The fact that $N^n$ is not a boundary is no longer relevant.
Clearly this is a simple and compelling way of dealing with non-boundary
manifolds; it is hard to see how else they can be handled. Furthermore,
we shall see below that the more familiar bulk/boundary formulation of AdS/CFT can be regarded as a limiting case of this geometry. In view of this, we advocate the hypersurface interpretation of ``infinity'' in {\it all} cases, even when $N^n$ {\it is} a boundary.

\ms

Now let us consider the geometry of the compact manifold $N^n$.  It is clear that, because the function $f$ introduced above is not unique, the metric $g^M$ on the bulk, $M^{n+1}$, does not induce a Riemannian structure on $N^n$, but rather a {\it conformal} structure.  We may therefore take it, without loss of generality [23], that the relevant metric on $N^n$ has {\it constant} scalar curvature.  It was pointed out in [24] that this constant {\it must not be negative}, since otherwise the conformal field theory at ``infinity'' will become unstable.  Now this constraint on $N^n$ has extraordinary consequences for the bulk, $M^{n+1}$.  It turns out [18] that if [a] the compact manifold ${\hat M}^{n+1}$ is constructed via the spin cobordism argument given above, [b] the bulk $M^{n+1}$ is orientable, [c] $N^n$ has positive or zero scalar curvature, and [d] the metric $g^M$ on the bulk has Ricci tensor Ric $\dis{(g^M) = -{n \over L^2} g^M}$, {\it then} $M^{n+1}$ {\it cannot be geodesically complete}.  This follows from the very remarkable work of Witten-Yau [19] and Cai-Galloway [20].  In simple examples, the incompleteness manifests itself as one or more regions where $g^M$ fails to be smooth.  In other words, {\it in this formulation of} AdS/CFT, {\it Randall-Sundrum style jumps are perfectly natural} $-$ indeed, they are inevitable.  Physically, this means that a fully general formulation of AdS/CFT necessarily entails the presence of branes in the bulk $-$ as foreseen in [2].  Mathematically, it means that, when constructing bulk solutions of Ric $\dis{(g^M) = -{n\over L^2} g^M}$, we are free of the (very restrictive) condition of smoothness, and this allows new and interesting geometries for the bulk.

\ms

The simplest possible cobordism from $N^n$ to itself is given by a cylinder.  Identifying the two copies, we obtain ${\hat M}^{n+1} = S^1 \times N^n$, where $S^1$ is a circle.  Even in this simplest possible case, there are many interesting geometric structures, so we shall confine our investigation to this case. Again, we stress that even if $N^n$ {\it is}
a boundary, there are definite advantages in constructing the bulk in
this way, since, as we shall see, it is easy to construct Einstein metrics
on the bulk even if the metric of $N^n$ is very complex or not known 
explicitly. By contrast, this could be very difficult in the usual bulk/boundary approach.

\ms

Parametrise the circle by $\theta$ running from $-\pi$ to $+\pi$, and think of ``infinity'' as residing at $\theta = 0$.  We shall look for metrics of the general form

$$g^M = A^2(\theta) [L^2 d\theta \otimes d\theta + B^2(\theta) g^N], \eqno{({\rm II}.1)}$$

\nt where $L$ is a constant, $g^N$ is a Riemannian metric on $N^n$, $B(\theta)$ is a positive function which does not tend to zero as $\theta$ tends to zero, and $A(\theta)$ is a positive function such that $1/A(\theta)$ has a first-order zero at $\theta = 0$.  Then $g^M$ will be a metric on the non-compact ``bulk'', $M^{n+1} = [S^1-\{0\}] \times N^n$, such that points in the hypersurface $\theta = 0$ are ``infinitely far'' from any point in the bulk.  As we have stressed, the compact manifold at infinity can have an otherwise arbitrary topology - it could be the $K3$ manifold, for example.

\ms

Our objective is to solve the bulk Einstein equation, Ric $\dis{(g^M) = -{n\over L^2} \; g^M}$.  In this we will be guided by the following lemma, the proof of which is straightforward and will be omitted.

\bs

\iii {\bf LEMMA:} /// With the above notation, $g^M$ is an Einstein metric only if $g^N$ is an Einstein metric.

\ss

\nt In view of the requirement that the scalar curvature of $N^n$ should not be negative if the CFT at ``infinity'' is to be stable, this means that we have two cases to consider.  The more interesting of the two is the Ricci-flat case, so we begin with it.

\bs

\nt {\bf III. RICCI-FLAT ``INFINITY''}
\ss
We now set $n=4$ and take $N^4$ to be Ricci-flat: thus $N^4$ could be a four-torus $T^4$ (in which case the compact five-dimensional manifold ${\hat M}^5$ is $T^5$), or a quotient of $T^4$ by a finite group, or $K3$, and so on.  As we know, there are {\it no} smooth solutions of Ric $\dis{(g^M) = -{n\over L^2} g^M}$ of the kind we require, and we must bear this in mind when solving the equation; we must do it ``piecewise''.

\ms

When $N^4$ is Ricci-flat, the solutions have the following {\it local} form:
$$A(\theta) = 1/(K-\theta), \; \; \; B(\theta) = J, \eqno{({\rm III}.1)}$$
\nt where $K$ and $J$ are constants.  The global solution has therefore to be patched together from ``pieces'' of this form, so that $A(\theta)$ and $B(\theta)$ are continuous functions on $S^1 - \{0\}$, that is, on the set $(-\pi, 0) \cup (0,\pi]$.  An interesting global solution of this kind is the following:
$$\eqalignno{ 
g^M 
&={1\over \theta^2}\left[ L^2 d\theta \otimes d\theta + (2\alpha-\pi)^2 g^{RF}\right], \; \theta \in (0, \alpha] \cup [-\alpha,0), &\cr
&={1\over (2\alpha-\theta)^2} \left[ L^2 d\theta \otimes d\theta + (2\alpha-\pi)^2 g^{RF}\right], \theta \in [\alpha, \pi], &({\rm III.2}) \cr
&= {1\over (2\alpha+\theta)^2} \left[ L^2 d\theta \otimes d\theta + (2\alpha-\pi)^2 g^{RF}\right], \theta \in (-\pi, -\alpha]. &\cr}$$

\nt Here $\alpha$ is a constant angle in the range $\dis{({\pi \over 2}, \pi)}$, and $g^{RF}$ is Ricci-flat.  Note that the introduction of the parameter $\alpha$ is the inevitable consequence of the ``singularity theorem'' discussed earlier : it describes the relative positions of the points on the circle where the metric fails to be smooth.

\ms

Clearly $g^M$ is well-defined everywhere on ${\hat M}^5 = S^1 \times N^4$, except at $\theta = 0$, which is ``infinity''.  However, $g^M$ does induce a conformal structure on  infinity, represented by the Ricci-flat metric $g^{RF}$.  Elsewhere, $g^M$ is smooth except at $\theta = \alpha$, $\theta = -\alpha$, and , perhaps a little less obviously, at $\theta = \pi$.  At $\theta = \pi$ there is a positive-tension brane which we interpret as the brane-world; the topology is that of $N^4$, and the metric induced by $g^M$ is precisely $g^{RF}$.  As we move away from the brane-world (in either direction), the sections $\theta =$ constant rapidly shrink, the induced metric being $g^{RF}$ multiplied by the factor $\dis{{(2\alpha - \pi)^2 \over (2\alpha \mp \theta)^2}}$, which is less than unity.  This continues until $\theta = \pm \alpha$ is reached, where the scaling factor reaches the value $\dis{{(2\alpha - \pi)^2 \over \alpha^2}}$; beyond this, the sections increase in size without limit as $\theta = 0$ is approached.  If $\alpha$ is assumed to be only slightly larger than $\dis{{\pi \over 2}}$, then $\dis{{(2\alpha - \pi)^2 \over \alpha^2}}$ is positive but very small.  The physical picture is then as follows.  We have a Randall-Sundrum brane-world at $\theta = \pi$, with the manifold rapidly shrinking away to either side.  The ``narrow'' region, however, proves to be the entrances of a wormhole with negative-tension branes at the throats; the other side of the wormhole is an infinite region of the kind familiar in AdS/CFT theory.  Thus, the Randall-Sundrum world can indeed arise naturally in an AdS/CFT geometry : {\it it lives deep inside a wormhole}.  The extent to which each ``world'' can influence the other is controlled by the parameter $\alpha$, which determines the width of the wormhole throats.

\ms

Let us examine the ways in which this global structure addresses the questions raised in the Introduction.  First, the geometry does clearly reconcile the differing natures of the Randall-Sundrum and AdS/CFT geometries; the former arises within a wormhole in the latter; this explains the equality of the $L$ parameters on the two sides of the calculation in [10].  Second, the lack of smoothness in the Randall-Sundrum geometry is now {\it explained} by AdS/CFT - it is an inevitable consequence of the Witten/Yau/Cai/Galloway theorems.  Third, the topology of the brane world ($\theta = \pi$) is the same as that of AdS/CFT ``infinity'' ($\theta = 0$) : both are diffeomorphic to $N^4$, and both have essentially the same geometry (given by the Ricci-flat metric $g^{RF}$ on the brane world, and by the conformal structure represented by $g^{RF}$ at $\theta = 0$).  This is in contrast to the usual picture, in which ``infinity'' is $S^4$ and thus has a very different topology and geometry to the brane-world.  Thus, in the new picture, it is at least topologically possible to imagine a limiting process whereby the brane-world comes to resemble the AdS/CFT picture.  Before discussing that, however, we wish to point out the following simple yet very appealing property of the above geometry.  Notice that there is an obvious asymmetry between the AdS/CFT and the Randall-Sundrum sides of the wormhole : on one side the manifold flares out to ``infinity'', while on the other it begins to do so but is abruptly halted at the brane-world.  This happens simply because $\theta$ is bounded $-$ values beyond $\pi$ are meaningless.  Thus we can say that the $UV$ cutoff implicit in the brane-world {\it is a consequence of the topology of the five-dimensional world}.

\ms

Now let us ask what happens if $\alpha$ (which always lies in the range $\dis{{\pi \over 2} < \alpha < \pi)}$ is taken to be arbitrarily close to $\dis{{\pi \over 2}}$.  Note that the brane-world, at $\theta = \pi$, has a metric $g^{RF}$ which does not depend on $\alpha$, while the metric at $\theta = \alpha$ is $\dis{ {(2\alpha - \pi)^2 \over \alpha^2}} g^{RF}.$   For an observer at the throat, therefore, the brane-world tends to become infinitely large as $\alpha$ tends down to $\dis{{\pi \over 2}}$.  The distance $d(\theta, \pi)$ from any point at a fixed value of $\theta \in [\alpha, \pi]$ to the brane world is
$$\dis{d(\theta, \pi) = L \; \ell n \; \left( {2\alpha - \theta \over 2\alpha - \pi}\right)}, \eqno{({\rm III}.3)}$$

\nt and this becomes arbitrarily large; the brane world recedes to infinity.  In short, by taking $\alpha$ sufficiently close to $\dis{{\pi \over 2}}$, we can effectively (though not truly) isolate the brane-world from AdS/CFT ``infinity''; by taking it still closer to $\dis{{\pi \over 2}}$, we can cause the interior of the wormhole to resemble its exterior.  Thus we have a precise formulation of the notion that AdS/CFT can be recovered from the Randall-Sundrum cosmology by some continuous process.  The relevant parameter is essentially the width of the wormhole throats, or, if one prefers, the magnitude of the tension of the branes there (which is also related to $2\alpha - \pi$).

\ms

The reader is entitled to object that the metric (III.2) bears no obvious relation to the Randall-Sundrum metric (I.3) or to the Euclidean AdS metric (I.1).  Before we explain this, the reader may wish to reflect on the respective roles played by Schwarzschild coordinates and Kruskal-Szekeres coordinates in the study of the Schwarzschild solution in general relativity.  The former have their uses, but they are profoundly misleading regarding the true global structure of the underlying spacetime manifold [25].  In particular, they conceal the wormhole structure of the full spacetime.  We claim that the familiar coordinate $y$ does a similar disservice to the study of (I.1) and particularly (I.3).  To see the point, take (III.2) and change the coordinate $\theta$ as follows.  If $\theta \in [\alpha, \pi]$, define $y$ by
$$e^{y/L} = {2\alpha - \theta \over 2\alpha - \pi}, \; \; \; y \in [0,y_\alpha], \; y_\alpha = L \; \; \; \ell n \left( {\alpha \over 2\alpha - \pi}\right). \eqno{({\rm III}.4)}$$

\nt A straightforward calculation reveals that, for these values of $\theta$, the metric (III.2) becomes
$$\dis{g^M = dy \otimes dy + e^{-2y/L} \; g^{RF}}. \eqno{({\rm III}.5)}$$

\nt If $\theta \in (-\pi, -\alpha]$, define $y$ by
$$\dis{e^{-{y/L}} = {2\alpha + \theta \over 2\alpha - \pi}, \; \; y \in [-y_\alpha, 0)}. \eqno{({\rm III}.6)}$$

\nt In this range, (III.2) becomes
$$\dis{g^M = dy \otimes dy + e^{2y/L} g^{RF}}, \eqno{({\rm III}.7)}$$

\nt and so, for $y \in [-y_\alpha, + y_\alpha]$ we have
$$\dis{g^M = dy \otimes dy + e^{-2|y|/L} \; g^{RF}}, \eqno{({\rm III}.8)}$$

\nt and so we have recovered the Randall-Sundrum metric, except that the flat metric $g^F$ in (I.3) has been replaced by a Ricci-flat metric $g^{RF}$.  (The fact that this replacement does not affect the Einstein equation Ric $\dis{(g^M) = -{4\over L^2} g^M}$ is explained in [26], page 268.)

\ms

For $\theta \in (0,\alpha]$, define $y$ by
$$\theta = \left( {\alpha^2 \over 2\alpha - \pi}\right) e^{-{y/L}}, \ \ 
y \in [y_\alpha, \infty), \eqno{({\rm III}.9)}$$

\nt and for $\theta \in [-\alpha, 0)$ define it by
$$\theta = \left( {-\alpha^2 \over 2\alpha - \pi}\right) e^{+{y/L}}, \ \
y \in (-\infty, -y_\alpha]. \eqno{({\rm III}.10)}$$

\nt Then (III.2) becomes
$$g^M = dy \otimes dy + e^{+2|y|/L}\left( e^{-{4y_\alpha /L}} g^{RF}\right), \ \ 
y \in (-\infty, -y_\alpha] \cup [y_\alpha, \infty). \eqno{({\rm III}.11)}$$

\nt Bearing in mind the fact that $\dis{e^{-{4y_\alpha /L}} \; g^{RF}}$ is Ricci-flat since $g^{RF}$ is Ricci-flat, we essentially have here two copies of the (Ricci-flat generalisation of) the Euclidean AdS metric (I.1).

\ms

If we take the manifold $N^4$ to be the torus $T^4$, so that the five-dimensional world is $T^5$, and if we take $g^{RF}$ to be a flat metric on $T^4$, then (III.8) is {\it locally} identical to (I.3), and (apart from a constant scaling factor) (III.11) is {\it locally} identical to (I.1).  The true global structure of this space is, however, utterly unlike the picture suggested by (III.8) and (III.11).  In particular, (III.11) very misleadingly suggests that ``infinity'' (at $y = -\infty$ and $y = +\infty$) is {\it disconnected}, which would lead directly to the paradox discussed by Witten and Yau [19].  In fact, of course, ``infinity'' is connected.  (Let $y \to \infty$ in (III.9) and let $y \to -\infty$ in (III.10).)

\ms

Our claim, then, is that the familiar $y$ coordinate in the Euclidean AdS metric (I.1) and in the Randall-Sundrum metric (I.3) is akin to Schwarzschild coordinates : very useful for many applications, but profoundly misleading as an indicator of global structure.  The analogue of Kruskal-Szekeres coordinates here is the circular coordinate $\theta$, and the global structure is to be read off from (III.2).  We have found that this global structure accounts for AdS/CFT/Randall-Sundrum complementarity in a way that answers all of the questions raised in the Introduction.

\bs

\nt{\bf IV. \ RICCI-POSITIVE ``INFINITY''}
\ss
The other potentially interesting solutions of Ric $\dis{(g^M) = -{4\over L^2} g^M}$ of the form (II.1) are obtained when $N^4$ is a compact Einstein manifold of positive scalar curvature.  In fact, the metric $g^P$ on $N^4$ must be normalised to satisfy
$${\rm Ric} (g^P) = {3\over L^2} \; g^P. \eqno{({\rm IV}.1)}$$

\nt As in the Ricci-flat case, we know that there are no geodesically complete solutions, so we must be prepared to patch a global solution together from local pieces.  A local solution of the form (II.1) is given by
$$A(\theta) = {1\over 2} {\rm cosec} \; (K - {\theta \over 2}), \ \ B(\theta) = 2 \cos (K - {\theta \over 2}), \eqno{({\rm IV}.2)}$$

\nt where $K$ is a constant, and so we can obtain a global solution on $(-\pi, 0) \cup (0,\pi]$ of the following kind :

$$\eqalignno{
g^M
&= {\rm cosec}^2\left({\theta\over 2}\right)\left[{1\over 4}L^2 \; d\theta \otimes d\theta + \cos^2\left({\theta \over 2}\right) g^P\right], \theta \in (0,\alpha] \cup [-\alpha,0) &\cr
&= {\rm cosec}^2\left( \alpha - {\theta \over 2}\right) \left[{1\over 4}L^2 \; d\theta \otimes d\theta + \cos^2 \left( \alpha - {\theta \over 2}\right)g^P\right], \theta \in [\alpha, \pi]
&{({\rm IV}.3)} \cr
&={\rm cosec}^2 \left( \alpha + {\theta \over 2}\right) \left[ {1\over 4}L^2 \; d\theta \otimes d\theta + \cos^2 \left( \alpha + {\theta \over 2}\right)g^P\right], \theta \in (-\pi,-\alpha], &\cr}$$

\nt where $\alpha$ is a constant angle in $\dis{({\pi \over 2}, \pi)}$.  This is a solution of Ric $\dis{(g^M) = -{4 \over L^2} g^M}$ provided $g^P$ satisfies (IV.1).  Clearly $g^M$ is well-defined and continuous everywhere in the ``bulk'', and it induces the conformal structure represented by $g^P$ on ``infinity'', which is at $\theta = 0$.

\ms

This solution shares many of the virtues of (III.2) : there is a brane-world at $\theta = \pi$, deep inside a wormhole with throats at $\theta = \pm \alpha$; the brane-world has the same topology as ``infinity''; as before, the characteristic Randall-Sundrum lack of smoothness is explained by AdS/CFT, via the Witten/Yau/Cai/Galloway theorems; and so on. In (IV.3), the metric at $\theta = \alpha$ is $\dis{\cot^2 ({\alpha \over 2})g^P}$, 
which tends to $g^P$ as $\alpha$ is taken closer to $\dis{{\pi \over 2}}$.
On the other hand, the metric induced on the brane-world is $\tan^2 (\alpha)g^P$, which increases without bound as $\alpha$ decreases towards $\dis{{\pi \over 2}}$. Relative to the length scales used by an observer 
on the brane-world, therefore, the throats of the wormhole shrink to zero
size in this limit; the distance from them to the brane-world tends to 
infinity; as before, the interior of the wormhole comes to resemble its 
exterior, in a clear-cut way; there is a clear sense in which AdS/CFT is a ``limit'' of the brane-world.

\ms

If we choose $N^n$ to be the four-sphere $S^4$ with its usual metric 
$g^S$ of constant curvature $1/L^2$, then (IV.3) has another very 
interesting limit, namely that in which $\alpha$ tends to $\pi$. In this
limit, the brane world shrinks to zero size, the negative-tension branes
coincide with the brane-world, and the wormhole pinches off. Surprisingly,
however, the metric becomes more rather than less smooth as this occurs.
To see this, for $\theta \in [\alpha, \pi]$, define a coordinate $x$ by
$$2 \tan (\alpha - {\theta \over 2}) = {L \over 2x_\alpha - x} - {2x_\alpha - x \over L}. \eqno{({\rm IV}.4)}$$

\nt Here $x \in [x_\pi, x_\alpha]$, where $x_\pi$ and $x_\alpha$ are the values of $x$ at $\theta = \pi, \alpha$ respectively.  Then (IV.3) becomes
$$g^M = {4 \over \left[ 1 - \left( {2x_\alpha - x \over L}\right)^2\right]^2} 
\left[ dx \otimes dx + \left( {2x_\alpha - x \over L}\right)^2 g^S\right]. \eqno{\rm{(IV}.5)}$$

\nt This represents a piece of hyperbolic space, $H^5$, corresponding to one side of the bulk adjacent to the brane-world.  For $\theta \in [0,\alpha]$, define $x$ by
$$2 \tan \left({\theta \over 2}\right) = {L \over x} - {x\over L}, \; \; x \in [x_\alpha, L], \eqno{({\rm IV}.6)}$$

\nt and now (IV.3) is
$$g^M = {4\over \left( 1-{x^2\over L^2}\right)^2} \left[ dx \otimes dx + {x^2\over L^2} g^S\right], \eqno{({\rm IV}.7)}$$

\nt representing another piece of $H^5$, in this case extending to ``infinity'' at $x=L$.  As before, these coordinates put (IV.3) into a more familiar form, at the cost of completely concealing the topology of the underlying five-dimensional space $-$ which, in this case, is that of $S^1 \times S^4$. Notice however that if $\alpha$ tends to $\pi$, then $x_\alpha$ tends to zero, meaning that (IV.7) is valid over the whole 
range of values of $x$ from zero to $L$, and so, as claimed, we obtain 
{\it all} of $H^5$, with its usual {\it complete} metric. Since the 
wormhole has pinched off, the sphere at infinity may now be regarded as 
a boundary rather than merely as a hypersurface. [To be precise, we now
have two copies of $H^5$ identified along their conformal boundaries.]
The usual bulk/boundary geometry is now seen as a limiting case of the
wormhole geometry. [Strictly speaking, we cannot let the wormhole actually
pinch off, since that would change the topology; however, we can take 
$\alpha$ so close to $\pi$ that, from a physical point of view, the
wormhole has pinched off for all practical purposes.]
\ms

We conclude this section with the observation that there are, of course, many more solutions of Ric $\dis{(g^M) = -{4\over L^2} g^M}$ compatible with the topology $S^1 \times N^4$.  For example, (IV.3) may be replaced by a metric having the form
$$\eqalignno{
g^M
&= {\rm cosec}^2(\theta)[L^2 \; d\theta \otimes d\theta + \cos^2(\theta) g^P], \; \theta \in (0,\alpha] \cup [-\alpha,0) &\cr
&= {\rm cosec}^2(2\alpha-\theta) [L^2 d\theta \otimes d\theta + \cos^2(2\alpha-\theta)g^P], \theta \in [\alpha, \beta] &\cr
&={\rm cosec}^2 (2\alpha - 2\beta +\theta) [L^2 d\theta \otimes d\theta + \cos^2(2\alpha - 2\beta+\theta)g^P], \; \theta \in [\beta,\pi] &({\rm IV}.8)\cr
&= {\rm cosec}^2(2\alpha+\theta)[L^2 d\theta \otimes d\theta + \cos^2 (2\alpha+\theta) g^P], \; \theta \in [-\beta, -\alpha] &\cr
&= {\rm cosec}^2(2\alpha-2\beta-\theta) [L^2 d\theta \otimes d\theta + \cos^2 (2\alpha-2\beta-\theta)g^P], \; \theta \in (-\pi,-\beta] &\cr}$$

\nt where $\alpha$ and $\beta$ are constant angles which must satisfy
$${\pi \over 4} < \alpha < {\pi \over 2}, \ \ {\pi \over 2} < \beta < 2\alpha. \eqno{({\rm IV}.9)}$$

\nt This metric corresponds to a wormhole with throats at $\theta = \pm \alpha$.  The throats can be made arbitrarily narrow by choosing $\alpha$ sufficiently close to $\dis{{\pi \over 2}}$.  The wormhole contains {\it two} brane-worlds, at $\theta = \pm \beta$; these can be made arbitrarily large by taking $\beta$ sufficiently close to $2\alpha$.  The two brane-worlds are separated by a further throat at $\theta = \pi$, where there is another negative-tension brane.  It is clear that one can construct arbitrarily complicated solutions in this way : the point to bear in mind is that, since it is {\it impossible} to construct a smooth solution, there is nothing unnatural about metrics like (IV.8).

\eject

\nt{\bf V. \ CONCLUSION.}
\ss
The familiar bulk/boundary formulation of AdS/CFT must obviously be generalised to accommodate manifolds (such as $K3$ ) which are not boundaries.  It was predicted in [2] that doing this will require the introduction of ``branes or stringy impurities of some kind'' into the bulk.  This indeed proves to be the case [18].

\ms

The objective of this work has been to argue that {\it Witten's ``stringy impurities'' include our Universe}.  We saw that, within a framework sufficiently general to allow for non-boundary manifolds, an Einstein metric on the bulk cannot be geodesically complete.  We are led to consider 5-dimensional compact manifolds with metrics built up from {\it pieces} of the form (II.1).  An example of an Einstein metric constructed so that AdS/CFT can be explored for $K3$ is given by (III.2).  We find that the generalised ``AdS'' space contains a wormhole, inside which there naturally arises a Randall-Sundrum type brane surrounded by a region with the familiar metric (III.8). Finally, we have seen that the wormhole approach
does {\it not} require us to jettison the standard bulk/boundary formulation of AdS/CFT. Instead, the latter is now seen to be a special
limiting case, the case in which the wormhole effectively pinches off,
destroying the branes, thereby restoring geodesic completeness and 
effectively converting the infinity submanifold to a boundary. [This 
pinching off is reminiscent of the constructions discussed in [27].]

\ms

We do not, of course, claim to have a realistic cosmological model here.  In order to construct such a model, one would have to begin by understanding the Lorentzian analogues of the Witten-Yau and Cai-Galloway results.  [For ideas on how this might be done, using techniques given in [28], see [29].]  It is not unreasonable to hope that, in the spirit of [2], our investigations may serve as a useful guide.  In particular, our results suggest :

\ss

\i [a]/ That the Randall-Sundrum coordinate $y$ may conceal considerable topological complexity in the underlying 5-dimensional manifold.  (See also [30].)

\ss

\i [b]/That wormholes in the AdS/CFT bulk may be important.  (See also [31].)

\ss

\i [c]/ That the AdS/CFT/Randall-Sundrum complementarity should be interpreted as evidence that AdS/CFT {\it explains} the existence of the brane-world - that is, of our Universe.

\eject

The Randall-Sundrum geometry strikes many theorists as somewhat contrived:
how could such a peculiar structure arise? The answer advanced here is 
that, if wormholes develop in the AdS/CFT bulk, then a structure like
the brane-world must develop inside the wormhole. On the other hand,
the picture we have developed here does have two principal drawbacks.
(1): One of the advantages of the ``second'' RS scenario is that, unlike
the first [32], it does not involve negative-tension branes. Here we
have reintroduced them, at the wormhole throats.
(2): We have assumed that the angle $\alpha$ is slightly larger than
$\dis{{\pi \over 2}}$. Why should that be so?

\ms

Regarding point (1): Any in-depth discussion of brane tensions must 
take into account the remarkable results of [33], which shows that 
correctly identifying brane tensions in the string context can be a
very subtle matter. Leaving those issues to one side, however, we feel
that Visser and Barcelo [34] have argued very persuasively that violations
of the familiar positivity conditions on energy densities and tensions
are virtually inevitable when scalar fields are present $-$ as they are
when one attempts to make brane-world models more realistic [35]. Indeed,
it is argued in [35] that Randall-Sundrum style ``kinks'' can be smoothed
if all tensions are positive [though it is questionable whether even this
can be achieved in a fully supersymmetric treatment [14]]. Since the kinks
in our model certainly cannot be smoothed, we take this as further 
evidence that negative-tension branes are unavoidable in our approach
$-$ as indeed one would expect in any theory involving wormholes [21].

\ms

In fact, it is very questionable whether the traditional aversion to 
negative tensions is really justified here. It is argued in [35] that
negative tension branes are acceptable if they lie on space-time 
boundaries generated by taking a quotient of some manifold by a 
symmetry group which has fixed points. In our picture, the negative
tension branes lie on the boundaries of {\it two} such spaces [one
containing the brane-world, the other containing AdS/CFT infinity]
which have been sewn together along their boundaries, so that the 
negative branes correspond to ``internal boundaries'' [27]. Therefore,
our negative branes are like those in [35]: ``...just part of a 
background, not something which can be dynamically created anywhere
in space''. There is however one difference: the negative tension branes
of [35] are at the ``end of the Universe'', so we cannot ask what happens
if some object passes through them in either direction. Our negative 
branes lie between two regions of the five-dimensional world, so for
us this is potentially very important. We return to this below.

\ms

Whether or not our negative branes are as innocuous as we claim, they
are in any case hidden in the throats of the wormhole; if the latter 
are sufficiently narrow and far away, then the negative branes are 
effectively inaccessible. This brings us to the second point raised
above, the fact that $\alpha$ is so close to $\dis{{\pi \over 2}}$.
We can show that this is indeed the case as follows. Equation (IV.1),
together with the fact that the Ricci tensor is invariant under constant
re-scalings of the metric, implies that 
$${\rm Ric} ({\rm tan}^2({\alpha}) g^P) = {3\over L^2}{\rm cot}^2({\alpha}) \; ({\rm tan}^2({\alpha})g^P). $$
Here ${\rm tan}^2({\alpha}) g^P$ is the metric induced on the brane world by the metric
(IV.3). Thus the cosmological constant of the brane-world is ${3\over L^2}{\rm cot}^2({\alpha})$ [unless it is {\it exactly} zero, as in Section III]. This is positive and very small,
as observations suggest, only if $\alpha$ is slightly larger than $\dis{{\pi \over 2}}$.

\ms

Since $\alpha$ is indeed so close to $\dis{{\pi \over 2}}$, the throats of the wormhole
are very narrow and very far away from the brane-world. As the Randall-Sundrum scenario
is, despite appearances to the contrary, entirely local [36], the existence of the wormhole
cannot be detected. For example, the presence of the wormhole will not affect the detailed
comparison made in [10], with which we began. [Note incidentally that [10] uses the {\it exact}
AdS/CFT relation $N^2 = {\pi}L^3/2G_5$, which is only consistent if the cutoff is taken
at a large value of $y$ in (I.1); as we have seen, this means, once again, that $\alpha$ is slightly larger than $\dis{{\pi \over 2}}$.] In other words, despite the fact that the
real structure of the space used in this work is very different to the one used in [5],[6],
[9], and [10], it leads to essentially the same results.

\ms

Can our approach ever lead to something quantitatively new? We believe that the answer
is yes, because we hope that it will eventually be possible to understand the value of
$\alpha$ in a {\it dynamical} way. That is, possibly $\alpha$ was {\it not} close to 
$\dis{{\pi \over 2}}$ in the early Universe $-$ indeed, the formula for the cosmological
constant, ${3\over L^2}{\rm cot}^2({\alpha})$, suggests that $\alpha$ was originally quite
small, if the Universe had an inflationary period [37]. In that case, in the early
Universe the throats of the wormhole were nearby and relatively large, so that access
to the AdS/CFT region was not impeded. The consequences of this might well have been
profound. To analyse them, we will need a good understanding of the dynamics of moving,
asymmetrical branes [38], as well as of the precise ways in which signals can be 
transmitted from one side of a negative brane to the other. This latter problem
will be particularly challenging, but very recent advances in the relevant mathematical
techniques [39] offer hope that the wormhole picture will eventually make some specific
predictions related to the early Universe. Even at this point one can see the kind of
influence the AdS/CFT parent universe can exercise on the brane-world, as follows. The
topology of the brane world must be related [via {\it cobordism}] to that of the AdS/CFT
world. The AdS manifold itself can be generalised to other negatively curved non-compact
spaces with other topologies, but many possible topologies are forbidden by stability
considerations [40]. Thus, the existence of the wormhole will directly constrain the
topology of the brane-world.

\ms

The overall picture presented here may therefore be described as follows. One can imagine
that ``spacetime foam'' in the AdS/CFT background could lead to the formation of a 
wormhole. Inside this wormhole, the Universe forms as a positive-tension ``kink''. At
this early stage it is strongly influenced by the parent universe, but the dynamics 
causes $\alpha$ to approach $\dis{{\pi \over 2}}$, effectively cutting off the Randall-
Sundrum space from the parent. The smallness of the observed cosmological constant
is now simply a reflection of the fact that the wormhole throats are far away and
very narrow. The only current traces of the AdS/CFT parent are to be found in relics
$-$ including cosmic topology $-$ of the very early Universe.

\ms
\ms

\nt{\bf REFERENCES}
\ms

\i [1]/ O. Aharony, S. S. Gubser, J. Maldacena, H. Ooguri, Y. Oz, Large $N$ Field Theories, String Theory and Gravity, Phys. Rept. {\bf 323} (2000) 183 [hep-th/9905111].

\i [2]/ E. Witten, Anti de Sitter Space and Holography, Adv. Theor. Math. Phys. {\bf 2} (1998) 253. [hep-th/9802150].

\i [3]/ L. Randall and R. Sundrum, An Alternative to Compactification, Phys. Rev. Lett. {\bf 83} (1999) 4690.  [hep-th/9906064]

\i [4]/ E. Witten, http://online.itp.ucsb.edu/online/${\rm susy}_-$c99/discussion/

\i [5]/ H. Verlinde, Holography and Compactification, Nucl. Phys. {\bf B580} (2000) 264. [hep-th/9906182].

\i [6]/ S. S. Gubser, AdS/CFT and Gravity. [hep-th/9912001].

\i [7]/ C. S. Chan, P. L. Paul, H. Verlinde, A Note on Warped String Compactification, Nucl. Phys. {\bf B581} (2000) 156. [hep-th/0003236].

\i [8]/S. Nojiri, S. D. Odintsov, S. Zerbini, Quantum (in)stabiliity of dilatonic AdS backgrounds and holographic renormalization group with gravity, Phys. Rev. {\bf D62} (2000) 064006 [hep-th/0001192].

\i [9]/ S. W. Hawking, T. Hertog, H. S. Reall, Brane New World, Phys. Rev. {\bf D62} (2000) 043501. [hep-th/0003052].

\i [10]/ M. J. Duff, J. T. Liu, Complementarity of the Maldacena and Randall-Sundrum Pictures. [hep-th/0003237].

\i [11]/ S. Nojiri, S. D. Odintsov, Brane world inflation induced by quantum effects, Phys. Lett {\bf B484} (2000) 119 [hep-th/0004097].

\i [12]/ S. Nojiri, O. Obregon, S. D. Odintsov, (Non-)singular brane-world cosmology induced by quantum effects in d5 dilatonic gravity [hep-th/0005127].

\i [13]/ L. Anchordoqui, C. Nu{\~n}ez, K. Olsen, Quantum Cosmology and AdS/CFT.  [hep-th/0007064].

\i [14]/ M. J. Duff, J. T. Liu, K. S. Stelle, A Supersymmetric Type IIB Randall-Sundrum Realization.  [hep-th/0007120].

\i [15]/ R. Akhoury, A Variational Principle for Radial Flows in Holographic Theories [hep-th/0007041].

\i [16]/ J. Maldacena, C. Nu{\~n}ez, Supergravity description of field theories on curved manifolds and a no go theorem [hep-th/0007018].

\i [17]/ J. W. Milnor and J. D. Stasheff, Characteristic Classes, Annals of Mathematics Studies 76, Princeton University Press, 1974.

\i [18]/ B. McInnes, AdS/CFT for Non-Boundary Manifolds, JHEP {\bf 0005} (2000) 025. [hep-th/0003291].

\i [19]/ E. Witten and S. T. Yau, Connectedness of the Boundary in the AdS/CFT Correspondence.  [hep-th/9910245].

\i [20]/ M. Cai and G. J. Galloway, Boundaries of Zero Scalar Curvature in the AdS/CFT Correspondence.  [hep-th/0003046].

\i [21]/ C. Barcelo, M. Visser, Brane Surgery : Energy Conditions, Traversable Wormholes, and Voids, Nucl. Phys. {\bf B584} (2000) 415.  [hep-th/0004022].

\i [22]/ H. B. Lawson and M. L. Michelsohn, Spin Geometry, Princeton Mathematical Series 38, Princeton University Press, 1989.

\i [23]/ R. Schoen, Conformal Deformation of a Riemannian Metric to Constant Scalar Curvature, J. Diff. Geom. {\bf 20} (1984) 479.

\i [24]/ N. Seiberg and E. Witten, The D1/D5 System and Singular CFT, JHEP {\bf 9904} (1999) 017 [hep-th/9903224].

\i [25]/ R. M. Wald, General Relativity, Chicago University Press, 1984.

\i [26]/ A. L. Besse, Einstein Manifolds, Springer-Verlag, 1987.

\i [27 / K. Kirklin, N. Turok, T. Wiseman, Singular Instantons Made Regular 
[hep-th/0005062]. 

\i [28]/ G. J. Galloway, K. Schleich, D. M. Witt, E. Woolgar, Topological Censorship and  Higher Genus Black Holes, Phys. Rev. {\bf D60} (1999) 104039 [gr-qc/9902061; see also hep-th/9912119.]

\i [29]/ E. Witten, http://online.itp.ucsb.edu/online/${\rm susy}_-{\rm c}99$/witten/

\i [30]/ H. Boschi-Filho, N. R. F. Braga, Quantum fields in anti de Sitter spacetime and degrees of freedom in the bulk/boundary correspondence [hep-th/0009039].

\i [31]/ S. Nojiri, S. D. Odintsov, K. E. Osetrin, Dilatonic quantum multi-brane-worlds [hep-th/0009059].

\i [32]/ L. Randall and R.Sundrum, A Large Mass Hierarchy from a Small Extra Dimension, Phys.Rev.Lett. 83 (1999) 3370 [hep-ph/9905221].

\i [33]/ M. Cvetic, M.J. Duff, James T. Liu, H. Lu, C.N. Pope, K.S. Stelle, Randall-Sundrum Brane Tensions [hep-th/0011167].

\i [34]/ M. Visser and C. Barcelo, Energy conditions and their cosmological implications [gr-qc/0001099]. 

\i [35]/ O. DeWolfe, D.Z. Freedman, S.S. Gubser, A. Karch, Modeling the fifth dimension with scalars and gravity,  Phys.Rev. D62 (2000) 046008 [hep-th/9909134].

\i [36]/ A. Karch, L. Randall, Locally Localized Gravity [hep-th/0011156].

\i [37]/ S.M. Carroll, TASI Lectures: Cosmology for String Theorists [hep-th/0011110].

\i [38]/ R.A. Battye, B. Carter, Generic junction conditions in brane-world scenarios [hep-th/0101061].

\i [39]/ P.B. Gilkey, K. Kirsten, D.V. Vassilevich, Heat trace asymptotics with transmittal boundary conditions and quantum brane-world scenario [hep-th/0101105].

\i [40]/ B. McInnes, Topologically Induced Instability in String Theory [hep-th/0101136].
 
\bye